\documentclass[amssymb,amsmath,showpacs,mathtools,maps,pra,floatfix]{revtex4}
\usepackage{color,graphicx,bm}
\usepackage[titletoc,title]{appendix}
\usepackage{overpic}
\usepackage{upgreek}
\usepackage{graphicx}

\def\beq{\begin{equation}}
\def\eeq{\end{equation}}

\def\eps{\epsilon}

\begin{document}

\title{Electrical Resistivity of Polycrystalline Graphene: Effect of Grain-Boundary-Induced Strain Fields
	

}
\author{S.E. Krasavin and V.A. Osipov}

\vspace{0.3 cm}

\affiliation{
Joint Institute for Nuclear Research,\\
Bogoliubov Laboratory of Theoretical Physics\\
141980 Dubna, Moscow region, Russia\\
e-mail: krasavin@theor.jinr.ru osipov@theor.jinr.ru, 
}


\begin{abstract}

We have revealed the decisive role of grain-boundary-induced strain fields in electron scattering in polycrystalline graphene. To this end, we have formulated the model based on Boltzmann transport theory which properly takes into account the microscopic structure of grain boundaries (GB) as a repeated sequence of heptagon-pentagon pairs. The effect of strain field is described within the deformation potential theory. For comparison, we consider the scattering due to electrostatic potential of charged grain boundary.  We show that at naturally low GB charges the deformation potential scattering dominates and leads to physically reasonable and, what is important, experimentally observable values of the electrical resistivity. It ranges from 0.1 to 10 k$\Omega $$\mu $m for different types of GBs with a size of 1 $\mu$m and has a strong dependence on misorientation angle. For low-angle highly charged GBs, two scattering mechanisms may compete. The resistivity increases markedly with decreasing GB size and reaches  values of 60  k$\Omega $$\mu $m and more. It is also very sensitive to the presence of irregularities modeled by embedding of partial disclination dipoles. With significant distortion, we found an increase in resistance by more than an order of magnitude, which is directly related to the destruction of diffraction on the GB. Our findings may be of interest both in the interpretation of experimental data and in the design of electronic devices based on poly- and nanocrystalline graphene.

\end{abstract}
\maketitle
 


\section{Introduction}

As is known, large-area films suitable for industrial application are usually polycrystalline consisting of a large number of randomly distributed single grains separated by grain boundaries (GBs). This holds for CVD grown graphene, which is considered as a promising material for nanoelectronics (for example when designing highly sensitive electro-biochemical devices~\cite{cummings}) and thermoelectrics~\cite{lehmann,sandon}. 
However, the effect of GBs on electronic transport properties is not yet well understood. 
Summary of the experimentally observed values of GB-induced resistivity ($\rho _{GB}$) by using various measurement techniques is given in Fig.2 of Ref.~\cite{isacss}. They varies over a wide range of values from 0.1 to 100 k$\Omega $ $\mu $m and depend on many additional factors such as distance from the charge neutrality point (adjustable by gate voltage), GB type, connectivity, width and some others.  Neither experimental~\cite{huang,yu,jaur,fei,ogawa,tsen} nor theoretical~\cite{yazi,fereira,roche,peres} studies have so far provided clear evidence clarifying the mechanism of electron scattering on GBs in graphene. 

Experimentally shown that GBs in graphene are n-doped due to localized electrons at pentagon-heptagon (5-7) pairs forming GB, while environment is p-doped. This should indicate that GBs act as electrical barriers to charge transport~\cite{tapaz,koepke} thus reducing conductivity. Theoretically, such mechanism of scattering across GBs has been studied in Refs.~\cite{yazi,fereira,roche,peres}. The problem is that high $\rho _{GB}$ values (1 k$\Omega $ $\mu $m and more) are not achievable when calculating carrier scattering on a weakly charged GBs (for 5-7 rings  it is equal to $e^*\sim 0.02$e according to~\cite{tamura}). 
Several  papers discuss the impact of graphene wrinkles~\cite{Clark}, GB's disorder~\cite{Vansco}, roughness and zig-zagness of extended GBs~\cite{majee} which make it possible to approach and in some cases even significantly exceed (see, e.g., Ref.~\cite{majee}) the expected range of values.  


It should be noted, however,  that one of the most natural mechanism of electron scattering due to GB-induced  strain fields has not yet been considered. 
Grain boundaries in graphene are formed by linear chains of pentagon-heptagon pairs or, equivalently, of 5-7 disclination dipoles, which are a source of additional mechanical stresses. This provides a new scattering channel for charged carriers: the GB-induced deformation potential scattering. The deformation potential is defined through the trace of strain tensor. Some time ago we suggested a model which takes into account the finiteness of the GB~\cite{jp1}. The basis for that model was the analogy between disclination dipoles and finite walls of edge dislocations. More specifically, wedge disclination dipoles simulate finite dislocation walls. This allowed us to describe the features of electron and phonon scattering in polycrystalline materials due to long-range strain fields~\cite{jp2}. Recently, we presented a general scheme that allows us to calculate strain fields in graphene caused by GB of any size and shape as a sum of strains of 5-7 disclination dipoles~\cite{kras1}. This approach has been applied to the analysis of heat transport in polycrystalline graphene~\cite{kras1} and allows us to consider any possible configurations of GBs including closed defects like the Stone-Walles~\cite{kras2}.

In this paper, we extend the model with explicitly included internal structure of GBs to the case of electronic scattering in graphene. It can be applied for analysis of both individual GBs of any finite length and  polycrystalline samples with the network of GBs. We consider a combination of two main sources of electron scattering: (a) deformation potential scattering and (b) electrostatic scattering due to charged GBs. The resistivity on GBs with different misorientation angles is calculated as a function of electron density. 
Our approach allows us to take into consideration the case of non-straight GBs with structural irregularities through the inclusion of additional partial disclination dipoles (PDDs). Calculations of the electrical conductivity are performed within the Boltzmann approach at room temperature. The expressions for relaxation times are derived in the first Born approximation.

\section{Model}

Let us consider a  GB of finite length in graphene as a periodic array of pentagon-heptagon pairs lined up along a line (see Fig.1). 
\begin{figure} [tbh] 
	\begin{center} 
		\includegraphics [width=10.5 cm]{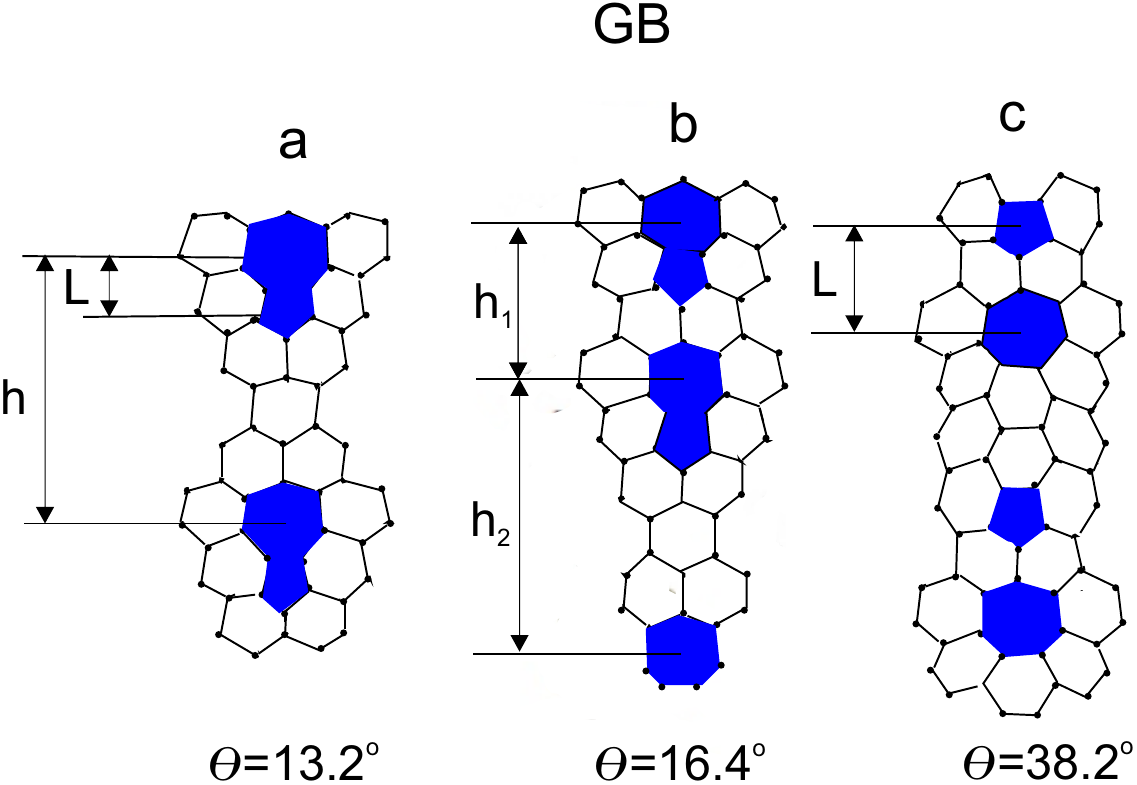}
	
	\end{center}
	\caption{Illustrative examples of GBs with different misorientation angles: (a) $\theta=13.2^{\circ }$, $L=0.246$ nm, $h=0.846$ nm; (b) $\theta=16.4^{\circ }$, $L=0.246$ nm, $h_1=0.42$ nm, $h_2=0.846$ nm; (c) $\theta=38.2^{\circ }$, $L=0.42$ nm, $h=1.01$ nm.}
\end{figure}

As is known, such array is a source of local stresses near the GB~\cite{romanov1}. Additionally, 
in the absence of a charged impurity, there is a small charge located on 5-7 pairs~\cite{tamura}. Therefore it is necessary to consider both possible mechanisms for the scattering of charged carriers: (a) strain fields caused by this defect which cannot be excluded, especially in the case of a low linear charge density at the GB and (b) electrostatic potential of the charged GB.

It is well known that the effect of strain field can be described within the deformation  potential theory~\cite{ziman}. In our approach, the total strain field caused by the GB at any point of graphene sheet is determined through a sum of strain fields from all 5-7 pairs~\cite{kras1}. In this way, for a GB oriented in the $x$-direction,
the deformation potential $V_{\epsilon}(r)$ takes the form  
\begin{equation}
V_{\epsilon}(r)=GTrE_{ij}(r)=G\frac {(1-\sigma )}{4\pi }\Bigl(\Omega \sum_{n=1}^{p}\ln\frac{(x-x_{n1})^2+(y-y_{n1})^2}{(x-x_{n2})^2+(y-y_{n2})^2}+\omega \ln\frac{(x-x ^{'}_{j1})^2+(y-y^{'}_{j1})^2}{(x-x^{'}_{m2})^2+(y-y^{'}_{m2})^2}\Bigr),
\end{equation}
where $TrE_{ij}(r)$ is the trace of the strain tensor, $G$ is the deformation potential constant, $\Omega $ is the modulus of the Frank vector which, for the chosen geometry, is directed along the $z$ axis, $p$ is the number of 5-7 pairs in the GB, $\sigma $ is the is Poisson's ratio, ($x_{ni}, y_{ni}$) are coordinates of $i$-th disclination in $n$-th dipole. In Eq.(1), the last term describes the dilatation for partial disclination dipole (PDD) with a power of $\omega $ located at points $(x^{'}_{m1(2)},y^{'}_{m1(2)})$ built in a GB. These  additional dipoles can appear  inside GBs because of the step-like variations in the misorientations along GB lines~\cite{ovid} or, in other words, disclinations forming the dipole are   points separating GB fragments with different tilt misorientation angles. If the dipoles are located along the $x$-axis, the coordinates $x_{ni}$ should satisfy the condition $|x_{n2} - x_{n1}|=L$, where $L$ is the length of the dipole arm in graphene.  

When determining the explicit form of the electrostatic potential $V_{Q}$ we consider that each charged 5-7 dipole is surrounded by a cloud of charges of opposite sign. In this approximation, 5-7 dipoles are localized point-like objects with a distance $\tilde{h}$ between them. The cloud radius $R$ lies in the region $L/2<R<R_{max}$ with $R_{max}=\tilde{h}/2$, that is a situation when neighbouring circles are touching each other. 
The Poisson equation takes the following form:
\begin{equation}
\nabla ^{2}V_{Q}(r,z)=-\frac{e^2}{\eps \eps_{0}}\Bigl[(N_{D}-N_{A})\sum_{l=0}^{p}\theta (R^2-(x-x_{l})^2-y^2)-\frac{e^{*}}{e} \delta(y)\sum_{l=0}^{p}\delta(x-x_{l})\Bigr]\delta(z),  
\end{equation}
where $r=(x,y)$, $N_{D}$ and $N_{A}$ are concentrations of donors and acceptors, respectively, $\theta (x)$ is the Heaviside unit step function; $x_{l}=x_{0}+l\tilde{h}$ is the coordinate of $l$-th dipole along the $x$ - axis, $e^{*}$ is the effective charge localized at 5-7 pair, and $\eps _{0} (\eps )$ is vacuum (relative) permittivity. 
Performing calculations by using the Fourier-transform method  we can find an expression for the effective two-dimensional electrostatic potential $V_{Q}(r)$
\begin{eqnarray}
V_{Q}(r)=\frac{e^2R}{4\eps \eps_{0}}(N_{A}-N_{D})\sum_{l=0}^{p} G_{22}^{11}\Bigl(\frac{R^{2}}{r_{l}^{2}}\Bigg\vert \  {{1}\;\;  {1} \atop \frac{1}{2} -\frac{1}{2}} \Bigr)-\frac{e^{*}e}{2\eps \eps_{0}}\sum_{l=0}^{p}\frac{1}{r_{l}},
\end{eqnarray}
where $r_{l}=\sqrt{(x-l\tilde{h})^2+y^2}$ and $G_{pq}^{mn}$ is the Meijer function~\cite{bateman}.

 Using two normalized chiral eigenstates $|{\bf k}>=\frac{1}{\sqrt{2}}{e^{-i\theta _{k}/2}\choose e^{i\theta _{k}/2}}e^{i\bf {k}\bf {r}}$~\cite{ando,peres1}, the scattering matrix for perturbation energy $V_{\epsilon}$ is found to be
\begin{equation}
<{\bf k}^{'}|V_{\epsilon }(r)|{\bf k}>=G\frac{(1-\sigma )}{2\pi }\cos\frac{\theta_{k^{'}k}}{2}\Bigl(-\frac{2}{q_{x}^2+q_{y}^2}\Bigr)\Bigl[\Omega \sum_{n=1}^{p}\Bigl(e^{i(q_{x}x_{n1}+q_{y}y_{n1})}-e^{i(q_{x}x_{n2}+q_{y}y_{n2})} \Bigr)+\omega \Bigl(e^{i(q_{x}x^{'}_{j1}+q_{y}y^{'}_{j1})}-e^{i(q_{x}y^{'}_{m2}+q_{y}y^{'}_{m2})}\Bigr)\Bigr],
\end{equation}
where ${\bf q}={\bf k}^{'}-{\bf k}$ and $\theta _{k^{'}k}=\theta _k^{'}-\theta _{k}$. By analogy, for the electrostatic energy given by Eq.(3) we get
\begin{equation}
<{\bf k}^{'}|V_{Q}(r)|{\bf k}>=\Bigl(e^2(N_{D}-N_{A})\frac{RJ_{1}(R\sqrt{q_{x}^2+q_{y}^2})}{q_{x} ^{2}+q_{y}^{2}}-\frac{ee^{*}}{(q_{x} ^{2}+q_{y}^{2})^{1/2}}\Bigr) \cos\frac{\theta_{k^{'}k}}{2}\frac{1}{2\epsilon \epsilon _{0}}\sum_{l=0}^{p}\exp(iq_{x}l\tilde{h}),
\end{equation}
where $J_{1}(z)$ is the Bessel function of the first kind.

In the framework of the Boltzmann approach, the conductivity in graphene is written as~\cite{novikov,vasko}
\begin{equation}
\sigma=\frac{4e^2}{h}\int_{0}^{\infty}EdE\Bigl[\frac{\tau _{+}(E)}{2\hbar }(-\frac{\partial f^{(0)}_{+} }{\partial  E})+\frac{\tau _{-} (E)}{2\hbar }(-\frac{\partial f^{(0)}_{-} }{\partial  E}) \Bigr],
\end{equation}
with 
\begin{equation}
n^{\pm}(\mu )=\int \frac{4d^{2}{\bf k}}{(2\pi \hbar )^2}f^{(0)}_{\pm }(E).
\end{equation}
Here $f^{(0)}_{\pm }(E)=1/[e^{(E \mp \mu)}+1]$ is the equilibrium Fermi-Dirac distribution function of electrons and holes with linear energy dependence $E=\hbar v_{F}|k|$, $v_{F}$ is the Fermi velocity, $n^{\pm}$ are the electron and hole densities, $\mu $ is the chemical potential measured relative to the half-filled $\pi $ band, $\tau _{\pm } (E) $ is the relaxation time for electrons and holes, and the factor $4$ accounts for spin and valley degeneracies. We consider the case when $n^{+}\approx N_{D}$ and  $n^{-}\approx N_{A}$.

In the first Born approximation, the relaxation times for the scattering mechanisms of interest to us are written as   \begin{equation}
\tau _{i}^{-1}(k) =\frac{n_{def}k}{2\pi \hbar ^{2}v_{F}}\int_{0}^{2\pi }d\theta |<{\bf k}^{'}|V_{i}(r)|{\bf k}>|^{2}(1-\cos \theta_{k^{'}k})
\end{equation}
where $n_{def} $ is the two-dimensional density of GBs, which can be easily determined for one-periodic structures if the distance $h$ between the dipoles is known (see Fig.1 (a,c)). Indeed, the parameter $h$ defines the GB size $D$ through the relation  $D=pL+(p-1)h$ with $p$ being the number of dipoles in the wall. The distance $h$ is directly related to the misorientation angle $\theta $, characterizing the type of GB (see, e.g., Ref.~\cite{romanov1}): the greater the $h$ the smaller the misorientation angle. For most of the GBs we examinated $D=1$ $\mu$m and $L=0.246$ nm, while $h$ changes. For example, for $\theta=9.4^{\circ }$ one has $h=1.27$ nm and $p=660$, for $\theta=21.8^{\circ}$ one has $h=0.42$ nm and $p=1500$ and the like. Two considered GBs are different: at $\theta=16.4^{\circ}$ (see Fig.1 (b)) two periods $h_1=0.42$ nm and $h_2=0.846$ nm occur (p=1140), while at $\theta=38.2^{\circ}$ (see Fig.1 (c)) the dipole arm increases to $L=0.42$ nm with $h=1.01$ nm ($p=695$). 
Evidently, the density of GBs is determined by means of the relation $n_{def}=1/D^2$, 
 so that $n_{def}=10^8$ cm$^{-2}$. The chosen value of $D$ allows us to compare our results with the existing experimental data in graphene with mesoscopic grain sizes (as, for example, in Ref.~\cite{tsen}). 
Notice, that it is also possible to prepare samples with poorly-connected GBs and for them $n_{def} $ will not be directly related to $D$. In the present work, we only focus on a case of fully-connected GB network. The relaxation time used in Eq.(6) is the combination of $\tau _{\epsilon }$ and $\tau _{Q}$ through the Matthiessen's rule~\cite{ziman}
\begin{equation}
\tau ^{-1}(k) = \tau ^{-1}_{Q}(k)+\tau ^{-1}_{\epsilon }(k). 
\end{equation}
Finally, the total resistivity $\rho_{GB}$ (defined as $\sigma^{-1}$) including the contribution from the two scattering mechanisms is calculated by means of Eq.(6).

\section{Results} 

We performed calculations of $\rho_{GB}$ at room temperature in a wide range of electron (hole) densities for GBs with different misorientation angles, both straight and non-straight (having additional disclinations). The results of our calculations are shown in Figs. 2-4. Let's start with the consideration of straight configurations.

\subsection{Straight GBs}

 First of all, let's make a few general conclusions. We found that the deformation potential scattering is the dominant scattering channel for all GB types provided that the total effective charge $e^{*}$ located at 5-7 pair is small. Indeed, its estimated value varies between 0.02 and 0.03 in units of electron charge $e$~\cite{tamura}. Our numerical results show that for $e^{*}=0.02e$ the ratio $\rho _{\epsilon}/\rho _{Q}$ takes values in the range from 5$\times10^{2}$ to 5$\times10^{3}$ depending on $\theta $. The highest values of $\rho _{\epsilon }/\rho _{Q}$ were found for denser GBs (with $\theta $ values around 32.2$^{\circ}$). We found also that $\rho _{Q}$ is slightly sensitive to the value of the screening parameter $R$. 

Fig.2 shows the resistivity of straight GBs  ($\omega =0$) as a function of $\delta n$ ($\delta n=n^{+}-n^{-}$) for different misorientation angles. 
\begin{figure} [tbh] 
	\begin{center} 
		\includegraphics [width=10.5 cm]{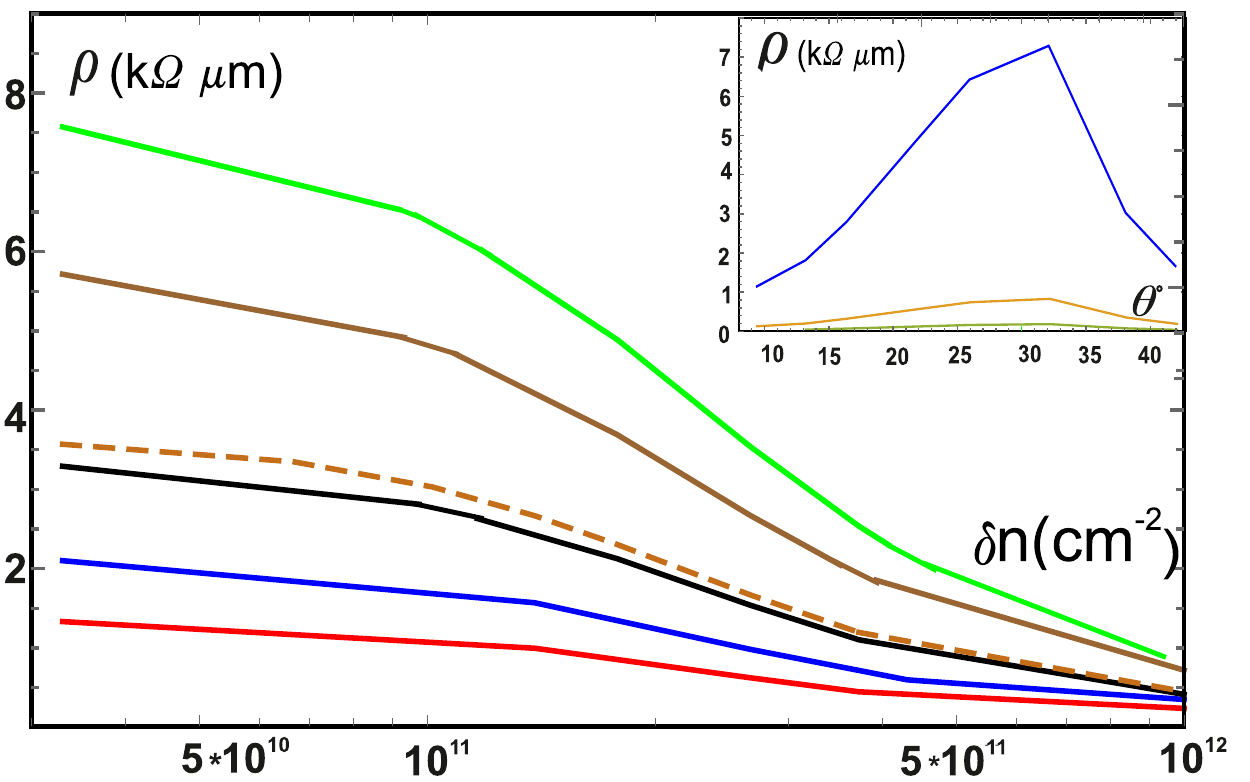}
	\end{center}
	\caption{Grain-boundary resistivity vs electron density at $T=300 K$ for different misorientation angles: 9.4$^{\circ }$ (red line), 13.2$^{\circ}$ (blue line), 16.4$^{\circ}$ (black line), 21.8$^{\circ}$ (brown line), 26.01$^{\circ}$ (green line), and 38.2$^{\circ}$ (short dashed line). The GB size is 1$\mu$m. Other parameters are: deformation-potential constant  $G=13$eV, effective electron charge at 5-7 dipole e$^{*}$=0.02e, $n_{def}$=10$^{8}$cm$^{-2}$, $\Omega$=60$^{\circ}$, $\sigma=0.2 $. The screening radius $R$ has greatest value for each GB. (Inset) Room-temperature GB resistivity as a function of misorientation angle $\theta ^{\circ}$ is shown in the inset at $\delta n=10^{11}$cm$^{-2}$(blue line), $10^{12}$cm$^{-2}$ (yellow line), and $3\times 10^{12}$cm$^{-2}$(green line).}
\end{figure}
As can be seen, $\rho _{GB} $ is growing with the misoriention angle up to $\theta\approx 32^{\circ}$ and takes the values from 1 to 7$ $ k$\Omega$ $\mu$m at  $\delta n\approx 10^{11}$cm$^{-2}$ .  It should be noted, however, that limitations of the applicability of the Born approximation do not allow us to accurately determine the value of the resistivity at low carrier densities (near the charge neutrality point $\delta n=0$). And yet, the behavior of the calculated curves gives grounds to assert that $\rho_{GB}$ reaches  values from 0.2 to 1.0$ $ k$\Omega$ $\mu$m, depending on the type of GBs. The values of $\rho_{GB}$ away from the electrical neutrality point are in good agreement with the experimental data of Ref.~\cite{tsen} if we use the relation between $\delta n$ and gate voltage $V_g$ in the form: $\delta n=\varepsilon^{\prime} V_g/(4\pi|e|d)$ with $\varepsilon^{\prime}$ being the dielectric constant of SiO$_2$ substrate and $d$ a distance from the back gate to the graphene sheet. It is important to note that $\rho_{GB}$ is sensitive to two model parameters $G$ and $\sigma$ through the factor $G^2(1-\sigma)^2$, whose values are not well defined. According to available estimates, $G$ can take values in the range from 7 to 19 eV, while $\sigma$ varies within 0.16 - 0.42. Respectively, the values of $\rho_{GB}$ can be both higher and lower than ours calculated at $G$=13 eV and $\sigma$=0.2. This remark is valid for all types of grain boundaries shown in Fig.2. 

The  biggest values of  $\rho _{GB} $ as well as of $\rho _{\epsilon }/\rho _{Q}$ relation have been found for GBs with angles from 26.01$^{\circ}$ to 32.2$^{\circ}$ where linear densities of 5-7 dipoles are maximal. 
A decrease in the density  of 5-7 dipoles also explains the decrease of $\rho_{GB}$ at $\delta n\approx 10^{11}-3\times 10^{12}$cm$^{-2}$ for $\theta$ above 32$^{\circ}$ (see the insert in Fig.2). Our calculations show that the smaller the GB wall size, the greater the resistance value due to an increase in the density of GBs. For example, for 21.8$^{\circ}$ GB with a size of $D=10$ nm we obtain $\rho _{GB}\sim 70$ k$\Omega$ $\mu $m at  $\delta n=10^{11}$cm$^{-2}$ in agreement with measurements in Ref.~\cite{kochat} (see also Fig.2 in Ref.~\cite{isacss}).  Notice that this value is more than an order of magnitude greater than that of the same GB with D=1 $\mu$m.

Let us briefly discuss under what conditions the importance of the electrostatic potential will increase. Obviously, this will require a significant increase in effective charge of 5-7 pairs, what can happen during doping of graphene by electron-donor and -acceptor molecules. In particular, it was shown that a single B or N impurity atoms prefer to incorporate into the grain boundary region and produce a p-type (n-type) doping in all investigated GB structures~\cite{brito}. Enhanced chemical reactivity of GBs allows one to consider polycrystalline graphene as a promising material for creating chemiresistors~\cite{Amin}, chemical ~\cite{yasaei} and biochemical~\cite{cummings} sensors and other applications. Fig.3 shows $\rho _{GB} $ as a function of $\delta n$ in the case when e$^{*}$ equals to 0.3e (green line) and  0.8e (red line). 

\begin{figure} [tbh] 
	\begin{center}
		\includegraphics [width=10.5 cm]{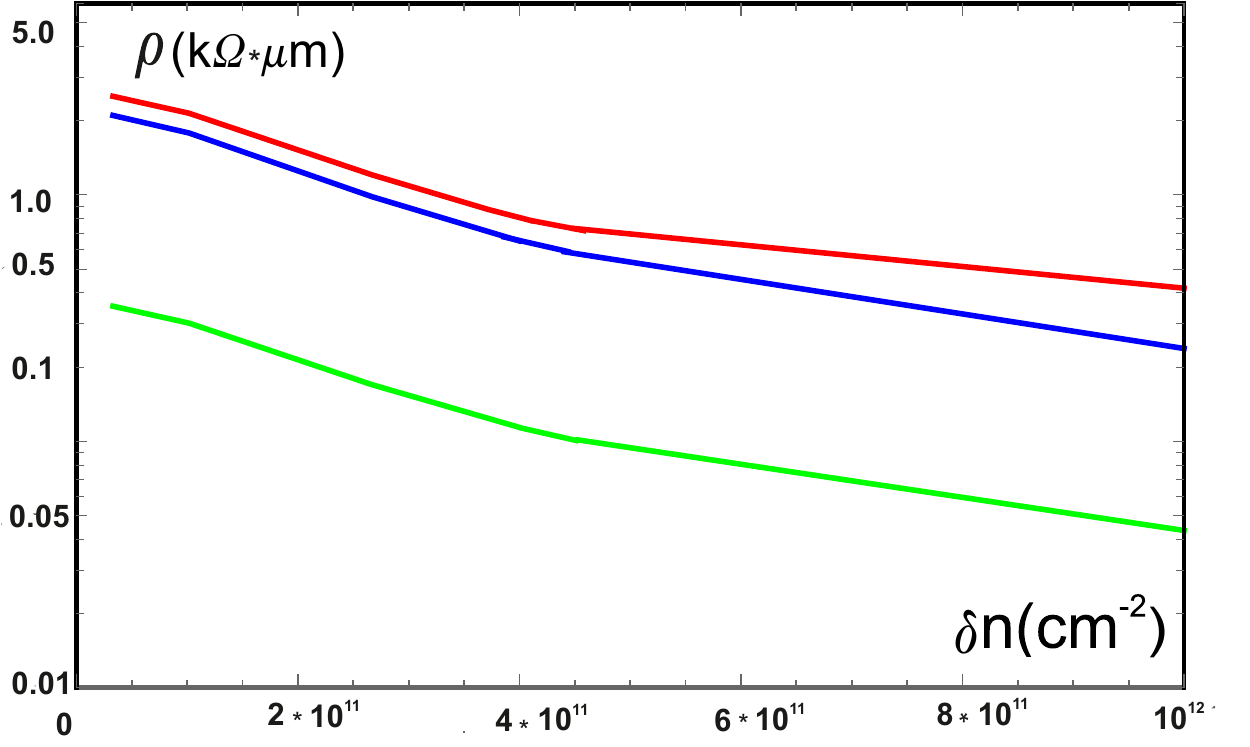}
	\end{center}
\caption{GB resistivity caused by electrostatic potential vs electron density at $T=300$ K for effective charges $e^{*}=0.3e$ (green line) and $e^{*}=0.8e$ (red line). The GB misorientation angle is taken to be  $\theta =13.2^{\circ}$ ($L=0.246$ nm, $h=0.846$ nm, $p$=916). For comparison, the resistivity due to deformation potential scattering from Fig.2 is shown by blue line.}
\end{figure}
For comparison, the contribution to $\rho _{GB} $ caused by the deformation potential scattering is given as well (blue line). We see that the scattering by electrostatic potential becomes comparable to the deformation-potential one when e$^{*}$ reaches the value of about 0.8$e$ and starts to  dominate at higher values. Thus, increasing effective charge on the wall leads to the higher resistivity values. This should be taken into account when analyzing experiments with charged GBs. Notice that the very possibility of competition between the two scattering mechanisms seems interesting to us. As can be seen from our calculations, this effect will be most pronounced at low-angle GBs.

\subsection{Non-straight GBs}

Let's take into account feasible irregularities in the spatial arrangement of 5-7 dipoles. Indeed, most of experimentally observed GBs in graphene sheets are curved and show no strict periodicity~\cite{ovid1}. This is especially true for walls with sizes of about a micron, which, as a rule, arise during synthesis by CVD method. A possible way to describe structural irregularities of real GBs in graphene is to embed partial disclination dipoles with strengths in the range of $-60^{\circ}<\omega<60^{\circ}$~\cite{ovid}. Earlier, in the study of heat transport in graphene (see Ref.~\cite{kras1}), 
we have shown the possibility, within the framework of our approach, of taking into account any number of PDDs  (including those with different $\omega$) inside the GB wall. Which is important in calculations, one can consider many built-in PDDs as one with a total arm length. 

Fig.4 shows the resistivity $\rho _{GB} $ as a function of $\delta n$ for 13.2$^{\circ}$ GBs containing a PDD with  the arms equal to $|x^{'}_{m2} - x^{'}_{j1}|=d=0.1$ $\mu$m (blue line) and $d=0.8$ $\mu$m (black line) ($\omega=45^{\circ}$).
\begin{figure} [tbh]
		\begin{center}

		\includegraphics [width=10.5 cm]{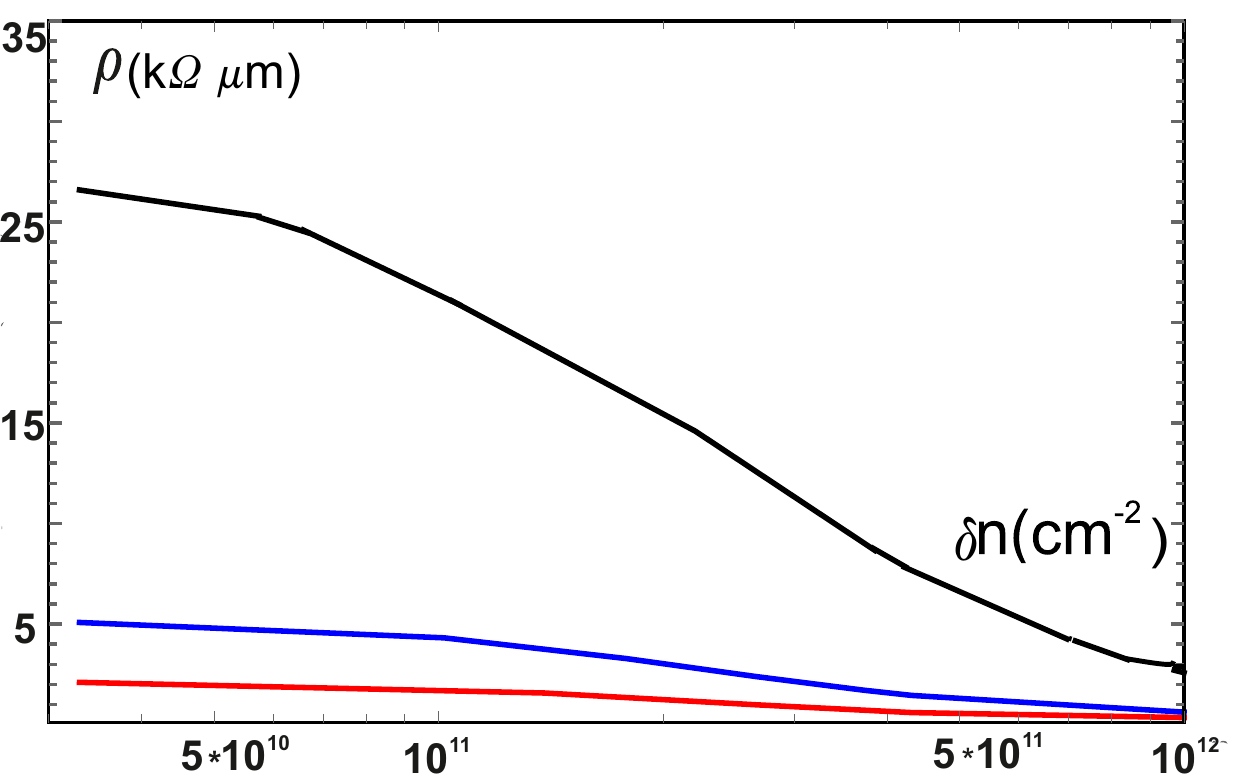}
	\end{center}
	\caption{GB resistivity as a function of electron density at $T=300$ K and $D=1 \mu$m in the presence of partial disclination dipoles of different arms: d=0.1 $\mu$m (blue line), d=0.8 $\mu$m (black line), and without a partial dipole (red line). The misorientation angle $\theta=13.2^{\circ}$, the strength of the partial dipole $\omega$=45$^{\circ}$. The deformation potential constant is equal to $13$ eV. }
\end{figure} 
As seen, for short PDDs there is a slight increase in resistivity (up to $\rho\approx 5.0$   k$\Omega$ $\mu $m near the charge neutrality point). With increasing size and/or number of regions with structural irregularities, the resistivity noticeably increases. In this case, the high values of $\rho _{GB} $ (2 k$\Omega$ $\mu $m and above) measured in some highly resistive samples (see, e.g., Refs.~\cite{jaur,tsen}) can be explained by deformation potential scattering only. We found that the more distorted the grain boundary, the less pronounced the dependence on the misorientation angle. This can also be seen from the long wavelength limit of the electronic mean free paths ratio $l_{GB+PDD}/l_{GB}\sim (1+\omega d/\Omega pL)^{-2}$. Here the dependence on $h$ disappears and, at fixed $\omega$, the effect is determined by the ratio of the total length of PDD arms $d$ and the effective length of 5-7 dipoles in the wall $pL$. Recall that $h$ is directly related to $\theta$. 

It is important to note that our analysis of the relaxation time behavior depending on the electron wavelength $\lambda $ 
 allowed us to draw an important conclusion about the specifics of electron scattering at grain boundaries in graphene. It turned out that this scattering behaves like in wave optics thus clearly manifesting the wave properties of electronic excitations in graphene. First of all, we found a strict proportionality of $\rho _{GB}$ to $p^2$ that is typical for Fraunhofer diffraction by amplitude gratings. At long wavelengths, the electron 'sees' a GB as a solid wall with size $D$. However, at small wavelengths, when $\lambda$ is compared to $D$ and further reduced, the GB becomes much more transparent due to its grating structure. As a result, the resistivity drops noticeably compared to what it would be for scattering on solid walls. Obviously, the larger $D$, the more pronounced the effect of reducing resistivity. This is exactly what happens when an additional PDD is embedded because it acts like a solid wall of size $d$ 'covering' the slots of the grate. Here the situation is reversed: the larger $d$ relative to $D$, the more significant the increase in resistivity (see also Fig.4).


%
\section{Conclusion}

In conclusion, in the framework of the proposed model based on Eqs.(1)-(3) we have demonstrated the possibility to reproduce the experimentally observed resistivities in polycrystalline graphene samples lying  between 0.1 and 100 k$\Omega$ $\mu $m for realistic parameters. Our study shows that the GB-induced deformation potential scattering gives the main contribution to the resistivity. Other important conclusions are as follows:

(i) in the case of straight GBs we found a strong correlation between resistivity and misorientation angle. The resistivity scales with grain diameter and can reach values of several tens of k$\Omega$ $\mu $m in nanocrystalline samples.  

(ii) in the presence of a noticeable charge on straight small angle GBs, an additional scattering channel becomes significant due to electrostatic potential. For GBs of mesoscopic size, the occurrence of competition between two scattering mechanisms is established.

(iii) at mesoscopic length, the GBs are usually not straight and contain structural irregularities. In this case, the deformation potential scattering increases noticeably due to additional strains caused by built-in partial disclination dipoles. This can lead to a marked increase in resistivity up to values of the order of 10 k$\Omega$ $\mu $m and more in agreement with experiments with highly resistive polycrystalline samples. Similar increase in resistivity in case of non-straight GBs was also reported in a recent paper~\cite{majee}.

As a final remark, our approach is quite universal and can be used to describe both electron and thermal resistivity in any polycrystalline 2D materials with GBs built from a sequence of 5-7 dipoles. Such a consideration is of obvious interest in the development of modern 2D materials with fundamentally new characteristics and the design of various electronic and thermionic devices since, as noted above, large-area films are usually polycrystalline.

\end{document}